\documentstyle[12pt,aaspp]{article}
\begin{document}

\def \GRBs{$\gamma$-ray bursts }
\def \KS{Kolmogorov-Smirnov }
\def \ks{{\it K-S }}
\def \vov{\left<V/V_{max}\right>}
\def \etal{{\it et al.}}
\def \be{\begin{equation}}
\def \ee{\end{equation}}

\title{Do Observations Exclude \GRBs Originating\\
in the Oort Cloud of Comets ?}

\vspace{0.2in}

\author{Eyal Maoz}
\affil{Harvard-Smithsonian Center for Astrophysics, \\
MS 51, 60 Garden Street, Cambridge, MA~02138}

\vspace{0.35in}
\centerline {$\dagger$ Accepted to publication in the
{\it Astrophysical Journal}}.

\vspace{0.15in}

\begin{abstract}
The currently favored explanation for the origin of \GRBs puts them
at cosmological distances;
but as long as there is no distance
indicator to these events all possible sources which are
isotropically distributed should remain under consideration. This is
why the Oort cloud of comets is kept on the list,
although there is no known mechanism for generating \GRBs
from cometary nuclei. Unlikely as it may seem, the possibility that \GRBs
originate in the solar cometary cloud
cannot be excluded until it is disproved.

We use the available data on the distribution of \GRBs (the BATSE
catalogue up to March, 1992), and
the Catalogue of Cometary Orbits  by Marsden and Williams (1992) to
investigate whether there is any observational indication for correlations
between the angular distributions of \GRBs and comets' aphelia,
assuming that the distribution of aphelia direction reflect,
at least to some extent, true variations
in the column density of the Oort cloud. We also apply the $\vov$
test to both distributions.

We have  performed a variety of statistical tests (a Kolmogorov-Smirnov
test for the distributions in galactic latitude, a
$\chi^2$ test for the spherical multiple moments, and a 2-D
cross-correlation analysis), including testing sub-samples for
isolating the effect of possible observational biases.
These tests imply that it is unlikely that the two distributions agree, but
the statistical significance  is not sufficient for ruling out any
connection with complete confidence. We performed Monte-Carlo simulations
which show that only when the number of bursts exceeds $\sim 800$
it is possible to rule out a correlation between the angular distributions.
Currently,
 it is only the combination of these tests  with the large disagreement
found for the $\vov$ parameter which makes the Oort
cloud of comets unlikely to be related to \GRBs.

\end{abstract}


\newpage
\section{INTRODUCTION}
Observations indicate that we are
in the center of a spherically symmetric distribution of \GRBs.
Currently, the favored explanation for the origin of these
bursts involves neutron stars at cosmological distances
(e.g., Paczy{\'n}ski 1992; Narayan, Paczy{\'n}ski, and Piran 1992,
and references therein) mainly due to the isotropy
of the observable universe on very large scales. However,
as long as there is no distance indicator for these
events, all possible astrophysical objects that are
isotropically distributed around us should remain under consideration.
This is the reason why
the Oort cloud of comets (Oort 1950; Weissman 1990) is kept on the list,
although there has never been a specific suggestion how to generate \GRBs
from cometary nuclei in the Oort cloud. In this paper we shall use the
available observational data on
both \GRBs and comets, and perform various statistical
tests in attempt to find
whether a relation between \GRBs and
the Oort cloud can be excluded, or alternatively, check whether there
is a statistically
significant correlation between these two populations.

If \GRBs originate in the Oort cloud their
number density in a certain direction should be roughly
proportional either to the
column number density of cometary nuclei in the cloud in that direction, or
to the projected square of three-dimensional density (if
collisions between comets is the underlying mechanism). Unfortunately,
we are unable to directly observe the Oort cloud itself but only
that tiny fraction of its comet population which happen to have a small
perihelia distance at their present passage near the sun.
We shall {\it assume} that variations in the flux
of the observed comets as a function of aphelia direction
reflect, at least to some extent, true variations in the column
density of the Oort cloud.

Such an assumption is crucial for being able to discuss deviations of the
cloud's structure from a perfect spherical symmetry. Since this central
assumption, although reasonable, cannot be verified observationally (but
see \S{5})
let us discuss some of its aspects. First,
the picture of a spherical comet cloud surrounding the planetary system and
stretching halfway to the nearest stars is only a rough approximation.
The aphelia directions of long-period comets, calculated from their original
orbits (before entering the planetary system) do not seem to be
distributed at random on the celestial sphere.
Variations in the flux of incoming comets from
different directions are evident.  For example, Oja (1975)
pointed to an overdensity of comet perihelia in a
direction close to the solar apex, and a concentration of comet
perihelia towards a plane which is inclined at 20 degrees to the
galactic plane, and Delsemme (1987) drew the attention to a deficiency
of comets near the galactic poles and equator.
Thus, the flux of the observed comets is already known to vary
with direction (see also Fig.1), and the coincidences of the detected
anisotropies with directions which are
of special interest imply that these may well
reflect real features of anisotropy rather than statistical fluctuations.

Our main concern is the possibility that the currently observed
variations in  the
distribution of comets' aphelia are attributed to some recent
perturbation(s) on the cloud.
It is indeed possible that the major features
were caused by recent
encounters with stars or molecular clouds, but Delsemme
(1986, 1987) has shown that in order to reproduce the observed variations,
an extremely slow
perturbing body is needed, moving typically at a relative
velocity of $200{\rm ms}^{-1}$.
Encounters with typical stars or clouds are very unlikely to
account for such effect, but a recent slow passage of a brown dwarf,
such as one of the many which may constitute the dark matter in the galactic
disk, is not improbable (Delsemme 1986).
 An encounter with a giant molecular cloud may in principle affect the
comet orbits dramatically, but the absence of a strong anisotropy feature in
the observed comet distribution, and the fact that
such encounters occur only about once
every 70-500 myr (depends on the impact parameter of the encounter)
make it very unlikely that we are currently experiencing a comet shower.
In fact, the smooth variation of the galactic tidal
field, rather than individual perturbations, is the dominant
mechanism which drives the evolution of the Oort cloud (e.g., Heisler,
Tremaine, and Alcock 1987), and the imprints of the galactic tides are
even claimed to be detected in the data (Delsemme 1987).
The tidal field sets limits on the outer dimensions of the cloud
and shapes it as a prolate spheroid with the long axis oriented
along the radius vector to the galactic center, which implies
that the column density
of comets in the cloud need not necessarily be identical in all directions.

Thus, it is reasonable to expect that anisotropies in the flux of the observed
comets are correlated, at least to some extent, to anisotropies in the column
density in the cloud. Still, this assumption is not secure
(see discussion in \S{5}). In \S{2} we describe the data, and in \S{3} the
statistical tests and their results. In \S{4} we compare the $\vov$
parameter of both distributions, and in \S{5} discuss
some caveats and summarize the main points.

\section{THE DATA}
The data on cometary orbits was kindly provided by Brian
Marsden from the Catalogue of Cometary Orbits (Marsden and
Williams 1992), with an addition of the few most recently
discovered comets. Short-period comets cannot have anything to do with
\GRBs since their distribution is highly anisotropic.
Therefore, our main
sample (hereafter denoted by $C_0$) includes only those comets
which satisfy the two following conditions: a) an orbital period
greater than 200 years, which is the traditional definition of
a long-period comet. b) classified in the catalogue as class 1 or
2 (the classification scheme is described by Marsden \etal\/ 1978),
namely, being widely observed and having a well determined orbit.
This last condition is required in order to
exclude comets with poorly determined orbits since those may be
erroneously classified as long-period ones, and
to avoid including poorly observed comets, such as very faint ones,
which are likely to introduce selection biases (see discussion below).
Finally, the trio 1992II, 1963V and 1965VIII, and the pair
1987XXX and 1988III, were each regarded as a single comet since
their identical orbital parameters indicate that
they separated from each other at some earlier perihelion passage.
Thus, the sample $C_0$ contains 272 long-period comets with good
orbit determinations and with no obvious trouble with
non-gravitational forces.

Clearly, there are comets whose discovery has been missed, but
this need not concern us since we are interested in the relative
fluxes of comets incoming from different directions. However, the
discovery probability of a comet depends on the number of observers
that have a chance to detect it, so there might be an
observational bias due to the different geographic
distribution of observers in the
northern and southern hemispheres. Although
comets which are visible in the south are likely to be
also visible sooner or later in the north, the fact that comets
become brightest near perihelion implies that there  may
still be a smaller number of cometary perihelia discovered in the southern
hemisphere which is an artifact of the observers' depletion.
Indeed, during the years
1840-1919 the number of observers in the southern hemisphere was
quite low, but the
situation has much improved after 1920 (Everhart 1967; Marsden 1992, private
communication), so since then this selection effect is
likely to be very small (the discovery probability
is not linear with the number of observers, but
converges rapidly when the observers' number approaches a certain
value).
Therefore, in order to isolate the effect of such possible
observational bias we shall examine also a subset of $C_0$,
hereafter denoted by $C_1$, which contains only those 192 comets that
have been discovered during or after 1920.
Indeed, 102 comets ($53\%$) in $C_1$ have perihelia directions in the Northern
hemisphere, which implies that a strong bias is very unlikely.
A comet with a large perihelion distance will always be
faint and thus have a smaller discovery probability (Evhart 1967), but there
is no obvious reason how this can introduce a bias in the distribution of
aphelia directions.

It is also possible that some of the observed long--period comets have
already completed several passages through the planetary system and thus may
have changed their original aphelia direction due to planetary perturbations.
Although our
samples may be contaminated by such comets, this is unlikely to
be a severe problem. Simulations show that originally long-period
comet that enter the inner planetary system do not survive for many periods
in this state but are either captured more strongly or ejected from the
solar system (e.g., Fernandez and Jockers, and references therein).
This is also evident from the substantial difference between
the "original" orbit parameters and the "future" ones calculated for the
observed long-period comets (Marsden and Sekanina 1973).
Therefore, we do not expect a substantial mixing of
aphelia directions for the observed long-period comets.
Most of them
are probably making their first passage through the inner part of the
solar system (Marsden, Sekanina, and Everhart 1978), and most of the others
are very unlikely to have considerably changed
their aphelia direction, the more so for comets with larger aphelia distances.
Still, we shall consider also a third
sample of comets, hereafter denoted by $C_2$, which is a subset of $C_1$
and contains only the {\it very\/} long-period comets, i.e., only those $96$
comets with $(1/a)$ smaller than the median value
in the sample $C_1$.

The data on \GRBs is based on the BATSE catalogue and includes the 260
triggered
bursts observed from April, 1991  until March, 1992 (this data set is hereafter
denoted by $B_0$). In order to examine possible
correlations between {\it weak\/} bursts and
{\it very\/} long-period comets
we have constructed also a subset of $B_0$ (hereafter denoted by
$B_1$) which includes only those 130 bursts with peak flux rate (on the 64ms
time scale) below the median value.

\section{STATISTICAL TESTS}
In this section we study the possibility that the frequency of \GRBs is
proportional to the number density of cometary nuclei.
This is an adequate assumption for cases in which bursts
are produced by an (inconceivable)
mechanism which does not involve cometary collisions (e.g., collisions with
other objects or internal processes). We shall
discuss a possible relation to the
square of the comet number density in \S{4}.

We thus wish to obtain an answer to the following question: is there any
observational indication that the directions of \GRBs and comets' aphelia are
drawn from the same underlying distribution? Alternatively, can we disprove, to
a certain significance level, the null hypothesis that the two data sets are
consistent with a single distribution function?
We shall perform three types of tests: \ks ones, a $\chi^2$ test for the
spherical multiple moments of both distributions, and a 2-D cross-correlation
test.
\subsection{\KS tests}
The \ks test is based on measuring the difference between two cumulative
distributions of one-dimensional data sets, but a {\it cumulative\/}
distribution is not well-defined for a 2-D set. Although the \ks test can be
generalized to distributions in a plane  (Press, \etal\/ 1992) it becomes
completely meaningless for distributions on a sphere where the two
coordinates are cyclical.
We  shall apply the familiar \ks test to the
1-D distributions of the data points in {\it galactic latitude}.
This has the advantage of not involving any binning in
galactic latitudes,
but some information is still 'thrown away' due to ignoring
the distribution in $l$. In general, such a test
would not be of much interest due to the complete freedom in
choosing the orientation of the spherical coordinate system. But
in our case, galactic tides are expected to affect the shape of the
Oort cloud in a way which is likely to be symmetric with
respect to rotations around the galactic $\hbox{\^z}$ direction. Thus, a
\ks test for the distributions in {\it galactic\/} latitude
is of special interest.

We performed such \ks test for each of the six possible
combinations (the $B_0$ and
$B_1$ sets versus $C_0$, $C_1$, and $C_2$ (see definitions in \S{2})).
The "\ks" columns in Table 1 show
the significance level for the null hypothesis that the two data sets
were drawn from the same distribution. We see that weak bursts turn to
have higher probability for correlation with the comet distribution, but
these results do not give a firm answer to whether the
differences between the distributions are due to statistical
fluctuations or reflect real inconsistency (the values are not small enough
to imply a statistically significant difference).
The loss of information which is associated with reducing the spherical
data to 1-D
distributions is apparently too large. Let us see now   how these results
change in 2-D tests.

\subsection{A $\chi^2$ Test for Spherical Multiple Moments}
\def \cave{\hbox{\=c}}
\def \cavexi{\overline\xi}
If the two data sets are drawn from the same underlying distribution
we expect their dipole and quadrupole moments
to be similar up to statistical errors. Instead of comparing each of the three
dipoles and five quadrupoles separately, we shall
combine all comparisons into a single
$\chi^2$ test with eight degrees of freedom, one for each spherical
multiple moment.
To do that we need to have the error estimates
(or variance) for these quantities, but how can this be done without
knowing the true distribution?

A powerful technique for estimating the statistical errors is the
bootstrap (or resampling) method (Efron and Tibshirani 1986;
Press \etal 1992; Press, Rybicki, and
Schneider 1992). It
uses the actual dataset with its $n$ data points
to generate a large number of synthetic data sets,
each also with $n$ data
points. The procedure is simply to draw randomly $n$ data points at a time
with replacement from the actual set. Obviously, one gets sets in which
a random fraction of the original points, typically $\sim 1/e$ of them,
are replaced by duplicated original points. For each of the resampled
data sets we calculate the statistical quantities, in our case the
spherical multipole moments, exactly as we did for the actual data set. The
simulated measured quantities can be shown to be distributed around the
corresponding quantities from the actual set in the same
way, both variance and covariance, that the
last are distributed with respect to the true values.

We have calculated the first eight spherical
moments, $c_i$ and $b_i$ ($i=1,..,8$) of the comets and \GRBs distributions,
respectively, and generated a $1000$ repeated resampled sets for both
distributions. Denoting
the $k$-th determination of the $i$-th quantity by $c_i^k$ (and $b_i^k$),
$k= 1,...,N$, the standard 1-$\sigma$ errors, $\sigma(c_i)$, are
estimated by
\be \sigma^2(c_i) \approx {1\over N} \sum_{k=1}^{N}(c_i^k - \cave_i)^2 \ee
where ${\hbox{\=c}}_i \equiv {N^{-1}}\sum_{k=1}^N(c_i^k)$, and $N=1000$.
If the errors of
the various quantities had not been correlated the $\chi^2$ would
have been
\be \chi^2 = \sum_{i=1}^{8}{(c_i - b_i)^2 \over \sigma^{2}(c_i)+
\sigma^2(b_i)} \ee
However, although the multiples are orthogonal to each other,
the statistical errors of the various multipoles are not statistically
independent because
the same data points are used for estimating the variance of each
moment. The correlation matrix
among the $c_i$ is estimated by
\be \gamma_{ij} \approx {1 \over N}\sum_{k=1}^{N}( c_i^k - \cave_i )
(c_j^k - \cave_j)  \ee
and a similar matrix $\beta_{ij}$ for the correlation between the
$b_i$ quantities.
The $\chi^2$ test with 8 degrees of freedom (e.g., Barlow 1989) reads
\be \chi^2 = \sum_{i,j=1}^{8} (c_i - b_i) (\gamma_{ij} +
\beta_{ij})^{-1} (c_j - b_j)  \ee
where the summation of the two correlation matrices is due to the fact
that the variance of the difference between quantities equals to the sum
of the variance of each quantity.
This test was again performed for the six combinations of data sets, and
the results are displayed in the "$\chi^{2}$" columns of Table 1. Again, the
weak bursts show a slightly higher probability for agreement but these
significance levels are far from being conclusive.

\subsection{A Cross-Correlation Test}
A rapid visual assessment of the equal-area projections (Figures 1,2)
seems to suggest that both distributions are roughly isotropic,
but contain some slightly overdense and underdense regions.
We wish to find whether over-densities (and under-densities) in the
two distributions coincide, namely, whether there is a statistically
significant correlation of small-scale
anisotropies. In order to do that we shall
employ a cross-correlation technique combined with a $\chi^2$ test.

Investigating correlations between qualitative features in the data sets
requires having an
estimate for the true density giving rise to the data.
The key point concerning density estimation from a discrete set of
points is that it involves some degree of smoothing.
Essentially, the larger the data set, the less smoothing is required
(a smaller smoothing scale) so in the limit of an enormous data set we would
do no smoothing at all.
There is much freedom in determining the best angular smoothing scale, but
it should satisfy the following conditions: a)
be larger than
the average nearest-neighbor angular separation
between data points ($\sim{7^{\circ}}$ in our case) since otherwise we try to
associate specific bursts with specific comets. b) It should be
substantially smaller than $2\pi$  radians since otherwise most possible
features will be erased, and all distributions will look alike.
The test of the spherical multiple moments (\S{3.2})
essentially investigated correlations
between large scale anisotropies. Therefore, since we are now interested in
the small scale variations we shall choose a value which is closer to the
lower limit, e.g, a smoothing
angular scale of $20^{\circ}$. Enlarging the smoothing
scale even up to $60^{\circ}$ changed all the results (the $c\hbox{-}c$
columns in Table 1) by less than a factor of $3/2$.

We shall use the smoothing procedure of a spherical data described by
Fisher \etal\/ (1987), and Watson (1983). If a sample contains $n$
points on a surface of a sphere, each with polar coordinates
$(\theta_i,\phi_i)$, then the estimated density of points at a given
direction $(\theta,\phi)$ is given by
\be f(\theta,\phi)={1\over 4\pi n \Theta_{e}^{2} \sinh(\Theta_{e}^{2}
)}\sum_{i=1}^{n} \exp^{\cos\Psi_{i} /\Theta_{e}^{2}}
\ee
where $\Theta_e$ is the effective smoothing angle (in radians),
$\Psi_i$ is the angular separation between the considered direction and
that of the i-th data point, and $\cos\Psi_i$ is given by
\be \cos\Psi_i \equiv (\sin\theta_i  \cos\phi_i \sin\theta \cos\phi) +
(\sin\theta_i \sin\phi_i \sin\theta \sin\phi) + (\cos\theta_i
\cos\theta)
\ee
The weight given to each data point depends only on its angular distance
from $(\theta,\phi)$ and not on its direction, and the normalization
coefficient is such that $\int{f(\theta,\phi)\,
d\Omega}=1$ for {\it any\/} spherical distribution of $n$ points.
Denoting the
dimensionless deviation of the density from the average density by
$\delta f(\theta,\phi)\equiv(f-{\hbox{\=f}})/{\hbox{\=f}}$,
where ${\hbox{\=f}}=(4\pi)^{-1}$, the
amount of correlation between variations in the
bursts distribution, $f_b$ (Fig.4), and variations
in the comet aphelia distribution, $f_c$ (Fig.3), is given by
\be \xi_{bc} = \int{\delta{f_{b}}(\theta,\phi) \, \delta f_{c}(\theta,\phi)
\, \sin\theta d\theta \, d\phi}       \ee
If variations in the two distributions correlate then $\xi_{bc}$
should be positive, and if the two data sets were drawn from a single
distribution than we should obtain $\xi_{bc}=\xi_{bb}$ up to statistical
errors, where $\xi_{bb}$ is the correlation of $\delta{f_{b}}$ with itself.

In order to find wether $\xi_{bc}$ is statistically consistent with
$\xi_{bb}$ we need an estimate for the associated
statistical errors, $\sigma_{bc}$ and $\sigma_{bb}$, respectively.
These are calculated
using the re-sampling method which was described in detail in \S{3.2},
and are given by
   $\sigma^{2}_{bb} \approx N^{-1}\sum_{k=1}^{N}{
(\cavexi_{bb} - \xi_{bb}^{k})^{2}}$, and a similar expression for
$\sigma^{2}_{bc}$,
where the index $k$ stands for the k--th re--sampling,
and $\cavexi_{bb}$
is the average of $\xi_{bb}^{k} ,k=1,..,N$ $(N=100)$.
The $\chi^2$ test for one degree of freedom thus reads
\be
 \chi^2 = {(\xi_{bb} - \xi_{bc})^{2} \over \sigma^{2}_{bb} +
\sigma^{2}_{bc}  }
\ee
The results (the "$c\hbox{-}c$" columns in
Table 1) show confidence levels which range from
$4\%$ to $18\%$ for agreement. Enlarging the smoothing scale up to values of
$60^{\circ}$ did not yield confidence levels higher than $25\%$.

\section{THE $\vov$ PARAMETER}
The BATSE observations (Fishman et al. 1991; Meegan et al. 1992) show
that $\vov \approx 0.35$ which indicates that the sources of \GRBs
have a density distribution which falls with the distance. However, the
spatial distribution of the strong bursts is found to be
roughly constant ($\vov\approx
0.5$).  This provides a challenge for the Oort cloud
hypothesis since it is unclear why should comet dynamics, which
is dominated by the potential of a single point mass, lead
the comet population to a density distribution which has a roughly flat core.
Indeed, numerical simulations of the origin and evolution of the solar
system cometary cloud (Duncan, Quinn, and Tremaine 1987) show that the
density profile of comets in the range 3000-50000 AU goes roughly as
$r^{-3.5}$, which does not agree with the roughly uniform
distribution of the strong \GRBs. This problem is even more severe if
cometary collisions are assumed to
produce the bursts, since the implied logarithmic
slope of the square of the comets number density is $-7$.

Let us check whether the spatial distribution of comets, as implied by the
{\it observed\/} ones, does indeed deviate significantly from a constant
density. This need not necessarily reflect the exact density profile of
the entire cloud since we are strongly biased to observe comets with a small
perihelia distance, which is most likely to be correlated with the
aphelia distance. Denoting the number of observed comets with a
semi-major axis in the range $[a,a+da]$ by $N(a)$, the implied total
number of comets with such $a$ is proportional to $N(a)T(a)$, where
$T(a)$ is the orbital period (for example, if we observe the
same flux of short-period and long-period comets it means that there
are much more unobserved long-period comets than short-period ones).
The number of comets, with a given $a$, that can be found in any given
time at a distance between $[r,r+dr]$ from the sun, is proportional to
the relative period of time that those comets spend in that distance
range, namely, $\propto dr\,v_{a}^{-1}(r))T^{-1}(a)$, where $v_{a}(r)$ is the
radial velocity of a comet with a given $a$ at a distance $r$ from the
sun. Thus, it is easy to verify that the
spatial distribution of comets, as implied by observations, is given by

\be n(r) \propto {1\over r^2} \int_{r}^{a_{max}}{
     {N(a)\,da \over \sqrt{ r^{-1} - a^{-1} }}} \ee

In order to estimate $N(a)$ we have extracted from the $C_0$ data set all
those comets with $3000\le a\le 50000$, and used a \KS test to find the
best power law which models the data. The significance level for an
$a^{-0.5}$ fit was the highest ($71\%$) and dropped rapidly to few percents
(and
below) for changes in the logarithmic slope of $\pm0.2$. Substituting the
above expression for $N(a)$ in equation (4.1) and integrating numerically we
obtain a logarithmic slope for $n(r)$ which changes from $-1.5$ to $-3$
between 3000 to 50000AU.

We conclude that both numerical simulations and observations imply a
spatial structure for the Oort cloud which is highly non-homogeneous.
An origin of \GRBs in the Oort cloud would require a mechanism that
produces a very fine-tuned luminosity as a function of distance from the
sun. This, and especially a relevance of bursts' generation to
collisions between comets, are thus extremely unlikely.

\section{CONCLUSION}
This study does not discuss if and
how \GRBs can be produced in the Oort cloud of
comets, but inquires how probable is this possibility, based on observational
data. We have performed a variety of statistical tests for correlations
between the angular distribution of both populations, and examined their
$\vov$ parameters. The tests of the angular distributions imply that it
is unlikely that the two distributions agree, but the statistical
significance is not sufficient for ruling out any connection with
complete confidence. We performed Monte-Carlo simulations in which we
gradually increased the \GRBs population by adding randomly distributed
data points and found that, assuming that \GRBs are distributed
isotropically, only when their number gets around $800\hbox{-}900$
the significance
levels of the angular distribution tests will be around $1\%$ or below.
Currently,
it is the substantial disagreement of the $\vov$ parameters, especially
if cometary collisions are assumed to be relevant, which makes the
Oort cloud of comets extremely unlikely to be the origin of \GRBs.

It should be noticed that, although reasonable,
it is not really
clear how well angular variations in the observed distribution of
aphelia directions represent true variations in the column
density of comets in the Oort cloud. A way to check this would be to perform
numerical N-body simulations and check the correlation between true
anisotropies
and the "observed" ones, but this is left to a future study. Also,
the current number of observed comets in not large enough that statistical
fluctuations
in the comet flux will be negligible. Even in the absence of a recent
encounter with a perturbing star or a molecular cloud, some of the detected
anisotropies may result from statistical fluctuations. In any case, our main
conclusion remains due to the large discrepancy in $\vov$.

Finally, although it is likely to be a coincidence, we draw the attention
to the peculiar resemblance of qualitative features in the angular
distributions of \GRBs and comets' aphelia (figures 3,4) in the celestial
hemisphere defined by $180\le l\le 360$, where $l$ is the galactic
longitude. This coincidence
does not show up in the statistical analyses due to the
stronger lack of correlation in the other hemisphere.

I am grateful to Brian Marsden for providing me the catalogue of cometary
orbits and for the enlightening discussions. I wish also to thank
Bill Press and Ramesh Narayan for the stimulating discussions and comments
on the manuscript, Bohdan Paczy{\'n}ski for the interesting discussions,
 and John Dubinski for the technical advices.
This work was supported by the U.S. National Science Foundation, grant
PHY-91-06678.

\newpage
\vspace{0.5in}
{\bf TABLE 1}

{\it The significance level for agreement between the angular distributions
of comets' aphelia and \GRBs}

\def \chisq{\chi^{2}}
\def \ccor{$c-c$}
\begin{tabular}{l||r r r|r r r}
&\multicolumn{3}{c}{$B_0 \, (n=260)$}
&\multicolumn{3}{c}{$B_1 \, (n=130)$} \\ \cline{2-7}
&\ks &$\chisq$ &\ccor &\ks &$\chisq$ &\ccor  \\ \hline\hline
$C_0 \, (n=272)$ & 11\% & 4\% & 8\% & 56\% & 30\% & 18\% \\
$C_1 \, (n=192)$ &  8\% & 28\% & 10\% & 35\% & 64\% & 10\% \\
$C_2 \, (n=96) $ & 14\% & 53\% & 9\% & 38\% & 53\% & 4\% \\
\end{tabular}
\vspace{0.3in}

{\bf Table 1} - The significance level for correlations between the
distributions of \GRBs and comets' aphelia directions. $B_0$ - all
bursts, $B_1$ - only the weak bursts, $C_0$ - all comets, $C_1$ -
only those discovered after 1920, $C_2$ - only the very long-period
comets from $C_1$. $n$ is the number of data points in each sample
(see details in \S{2})

\vspace{1.0in}

{\bf FIGURE CAPTIONS}

{\bf Fig. 1}: An equal-area (Aitoff) projection of aphelia directions of
long-period comets on the entire celestial sphere in galactic coordinates.
(Squares -- discovered before 1920, Circles -- discovered after 1920,
Triangles -- very long-period comets discovered after 1920). The celestial
equator is also drawn.

{\bf Fig. 2}: An equal-area projection of the directions of the 260
\GRBs in galactic coordinates.  Triangles stand for weak bursts.

{\bf Fig. 3}: An equal-area projection of the smoothed distribution of
comets aphelia directions, based on the 192 comets (the $C_1$ set)
discovered during or after 1920, with an effective smoothing scale of
$20^\circ$ (see \S{4}). Darker areas reflect higher angular densities,
and the grey levels are linearly enhanced for an easy visual assessment
of qualitative features.

{\bf Fig. 4}: The same smoothing procedure as in figure 3, but for the
\GRBs distribution (the $B_0$ set). These smoothed projections enable an
 easy visual comparison between angular variations in the two distributions.
A quantitive analysis is described in \S{3.3}.


\newpage
\begin{references}
\reference Barlow, R.J. 1989, In {\it Statistics\/}, John Wiley \&
 sons, New York, p.150
\reference Heisler, J., Tremaine, S., and Alcock, C. 1987, Icarus,
70, 269
\reference Delsemme, A.H. 1986, in {\it The Galaxy and The Solar
System}, R.Smoluchovski, J.N. Bahcall, M.S. Matthews, eds., U.
Arizona Press, Tucson, pg.173
\reference Delsemme, A.H. 1987, A\&A, 187, 913
\reference Duncan, M., Quinn, T., and Tremaine, S. 1987,
Bull.Am.Astron.Soc, vol.19, No.2, pg.742
\reference Efron, B., and Tibshirani, R. 1986, In {\it Statistical
Science\/} vol.1, pp.54-77
\reference Everhart, E. 1967, AJ, 72, 716
\reference Fisher, N.I., Lewis, T., and Emblenton, B.J.J. 1987,
 In {\it Statistical Analysis of Spherical Data\/}, Cambridge University
 Press.
\reference Fishman, G. et al. 1991, presentation for BATSE at GRO Workshop
   in Annapolis
\reference Marsden, B.G., Sekanina, Z. 1973, AJ, 78, 1118
\reference Marsden, B.G., Sekanina, Z., and Everhart, E. 1978, AJ, 83, 64
\reference Marsden, B.G., and Williams, G.V. 1992, The {\it Catalogue
  of Cometary Orbits}, 7th edition, IAU Central Bureau for Astronomical
  Telegrams.
\reference Meegan, C.A., et al. 1992, Nature, 355, 143
\reference Narayan, R., Paczy{\'n}ski, B., and Piran, T. 1992, ApJ Lett,
  395, L83
\reference Oja, H. 1975, A\&A, 43, 317
\reference Oort, J.H. 1950, Bull.Astr.Inst.Neth, 11, 91
\reference Paczy{\'n}ski, B. 1991, ACTA Astronomica, 41, 257
\reference Press, W.H., Teukolsky, S.A., Vetterling, W.T., and Flannery,
B.P. 1992, In {\it Numerical Recipes in Fortran}, Second edition,
Cambridge University Press.
\reference Press, W.H., Rybicki, G.B., and Schneider, D.P. 1992, CFA preprint.
\reference Watson, G.S. 1983, In {\it Statistics On Spheres}, University
   of Arkansas Lecture Notes in Mathematical Sciences, Volume 6. New York:
   John Wily, pg 14.
\reference Weissman, P.R. 1990, Nature, 344, 825
\end{references}
\end{document}